# Blending machine learning and physics-based approaches for weather and climate: a typology

Benjamin J Shipway, Caroline Bain, David Walters, Ben B. B. Booth, Ian Boutle, Robin T. Clark, Katherine L. Hill, Elizabeth Kendon, Simon B. Vosper

Met Office, Exeter, United Kingdom

*Corresponding author*: Ben Shipway, ben.shipway@metoffice.gov.uk

ABSTRACT

The integration of machine learning (ML) with traditional physics-based models is reshaping the landscape of weather and climate prediction. On their own, ML-based and physics-based approaches each have significant benefits - but also challenges. Deploying both these approaches side by side has the potential to accelerate the pull through of emerging science in a trusted and practical way. But there are many choices that can be made to how we “blend” ML and established physics-based modelling systems to get the optimal benefits. This paper aims to provide a typology of blended modelling approaches and discusses some of the strategic benefits that come with them. It can be used not just to classify modelling systems, but also identify routes to gradual, incremental or wholesale development and implementation of new and emerging capabilities. These approaches provide a practical path to innovation by combining the speed and adaptability of machine learning with the robustness, trust, and interpretability of physics-based systems. By adopting a structured vocabulary and outlining the benefits and limitations of each approach, this framework supports informed decision-making and strategic planning, and can be used by the wider community to navigate the transition to next-generation prediction systems.

SIGNIFICANCE STATEMENT

Blended modelling offers a practical and strategic path to innovation in weather and climate prediction. By combining machine learning with physics-based systems, it enables faster, more accurate and more energy-efficient forecasts while preserving trust, transparency and compatibility with existing workflows. Yet the growing range of blending options has created ambiguity in language, choices and expectations. A framework, centred on a clear typology, helps define these approaches and supports incremental adoption of new technologies, allowing operational centres to meet rising demand for localised forecasts and increasing climate risks. As computational constraints tighten and ML capabilities expand, discussing different blended strategies provides a balanced roadmap that maintains scientific rigour and operational reliability. The typology has relevance beyond weather and climate.

## 1. Introduction

Weather and climate prediction have historically relied on physics-based numerical models, an approach first articulated by Vilhelm Bjerknes (Bjerknes, 1904) and later advanced by Lewis Fry Richardson (Richardson, 1922; Schultz & Lynch, 2022). Practical implementation became possible with the advent of electronic computing, leading to the first computer-based forecast (Charney et al., 1950) and the introduction of operational systems in the 1960s. Climate modeling followed similar principles, beginning with early general circulation models (Phillips, 1956) and pioneering work on radiative-convective equilibrium (Manabe & Wetherald, 1967). These milestones, enabled by decades of computational advances, underpin modern forecasting and climate research, delivering critical insights for safety, economic resilience, and policy.

These physics-based models remain indispensable for scientific understanding and operational reliability. However, their computational cost rises sharply with increasing resolution, larger ensembles, and increased model complexity, constraining our ability to fully capture small-scale processes such as convection and turbulence. To address these gaps, parameterizations are employed to approximate unresolved phenomena, but they introduce uncertainties that can affect both forecast accuracy and climate projections (Bauer et al., 2015). Nesting very high-resolution regional models (on the order of 1 km or finer) within global models can provide greater detail (e.g. Bush et al, 2023), but these limited-area models (LAMs) also incur substantial computational expense as resolution increases.

While physics-based models remain central for operational weather prediction and climate science, recent breakthroughs in machine learning (ML) introduce a fundamentally different approach with transformative potential. Machine learning methods promise orders-of-magnitude reductions in computational cost, the ability to generate large ensembles at minimal expense, and the prospect of faster update cycles and higher-fidelity forecasts (Lam et al., 2023; Shi et al., 2025; de Burgh-Day & Leeuwenburg, 2023). Yet these benefits come with new challenges. ML approaches raise concerns around trust, generalization, and integration into established workflows. Their dependence on historical data raises questions about limitations to robustness under climate change, extreme events, and data-sparse regions, while interpretability and transparency remain major considerations for operational adoption (Shi et al., 2025).

In response, the weather and climate community is increasingly exploring hybrid strategies that combine the strengths of physics-based and ML methods (Slater et al., 2023; Arcomano et al., 2022; Quarteroni et al., 2025). These approaches offer a practical, incremental path to innovation - enhancing forecast skill, agility, and sustainability while preserving trust and continuity in existing systems. Recent scientific advances in hybrid modeling show promise (Farchi et al., 2024; Kochkov et al., 2024), the emulation of model components (e.g., physical parametrizations) potentially enables greater accuracy (Morcrette et al., 2025) and blending separate physics-based and ML-based simulations within ensembles has been shown to improve forecast skill (Trotta et al., 2025). For climate, ML downscaling emulators have shown real potential for enhancing regional climate simulations (Kendon et al, 2025), providing more comprehensive sampling of uncertainty at local scales.

As interest in these methods grows and the community navigates a range of options, it is essential to establish a common language and understand the opportunities and implications of emerging approaches. Terms such as “hybrid” and “online learning” are often used inconsistently. For example, “hybrid” sometimes refers to any mix of machine learning (ML) and physics-based modelling and the specific nuances of what this means in practice is sometimes unclear.

In this article, we propose a clearer typology. We define “hybrid” as ML embedded within physics-based systems and use “blended” as a broader term encompassing hybrid, integrated, augmented, and independent ML as well as physics-based approaches.

Our goal is to clarify this landscape and provide a framework that supports communication and strategic decision-making. We outline a spectrum of blending opportunities - from independent physics-based to independent ML systems - with intermediate options that leverage interactions between the two (Fig. 1). This typology is intended to guide discussions, inform prioritizations and investment decisions for research and development, and support the development of forecasting systems that are operationally viable, scientifically credible, and computationally efficient. Furthermore, at a time when the computing landscape is radically changing, this framework can help organizations in defining their future technology requirements.

1

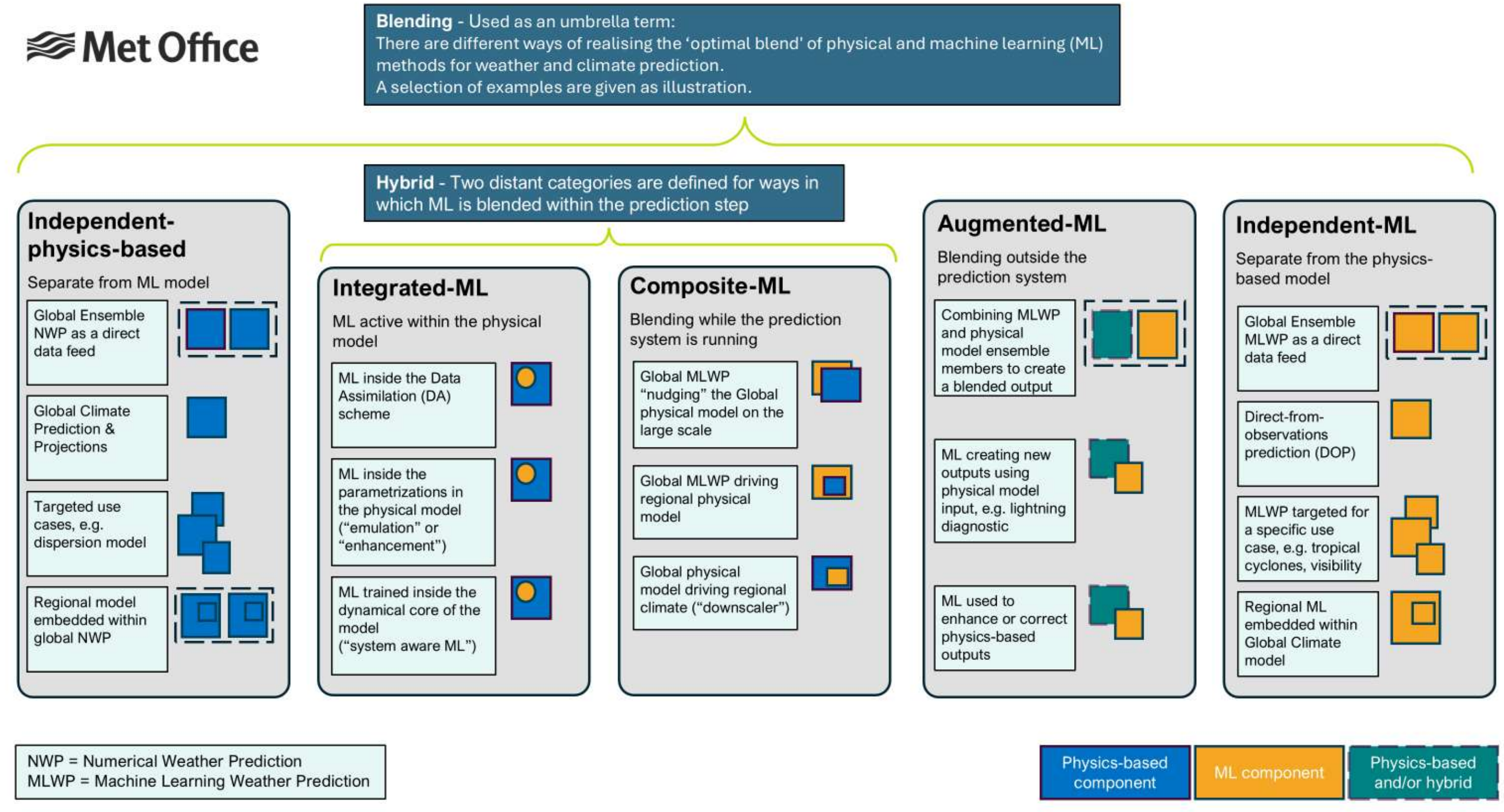

2

3 Fig. 1. We describe a spectrum of blending opportunities from independent-physics-based to independent-ML. Options in between make use
4 of the interaction between physics-based and ML in different ways.

## 2. Clarifying the Landscape

In discussion of the proposed typology it is useful to briefly review the components that may be deployed within an optimal blend:

*Physics-based models*

Weather and climate prediction have long relied on physics-based numerical models - such as General Circulation Models (GCMs) and Limited-Area Models (LAMs) - which simulate the Earth system using fundamental physical laws. These models solve the governing equations of fluid dynamics, thermodynamics, and related processes, combining representations of radiation, convection, cloud microphysics, and more to provide a physically consistent framework for forecasting and climate projection. They remain essential for advancing scientific understanding (Balaji et al., 2022) and operational reliability. Over the past decades, international efforts have progressively advanced these systems, leveraging exascale supercomputing architectures (Adams et al., 2019, Ubbiali, S et al, 2025), improved numerical schemes (Melvin et al., 2024, Zängl, G. et al., 2015, Skamarock, W.C. & Klemp, J.B. 2008), improved physical representation (e.g. Field et al. 2023, Jadav et al., 2025), and multi-resolution techniques (Brown et al., 2024, Herrington et al.,2019 ) to enhance fidelity and reduce uncertainty. Both in climate and in weather prediction systems, there is an increasing trend to couple separate models representing different components of the Earth System (Lewis et al., 2019, Dunne et al., 2020). This incremental development and the steady progression of forecast skill it has delivered has been described as a "quiet revolution" with impact among the greatest of any area of physical science (Bauer et al., 2015).

However, physics-based models face inherent challenges. Their computational cost grows steeply with resolution, limiting the ability to fully resolve small-scale phenomena such as convection or turbulence. Parameterizations, which approximate these unresolved processes, introduce uncertainties that can affect forecast accuracy and climate projections. While these models are robust and interpretable, their complexity and resource demands further drive interest in complementary approaches.

*Machine Learning models*

Machine Learning Weather Prediction (MLWP) models have emerged as one of the most disruptive but opportune developments in weather prediction in decades. Unlike physics-based models, which explicitly solve physical equations, MLWP systems learn statistical relationships from data. The rapid advances in deep learning, combined with the availability of massive reanalysis datasets and high-performance computing, have enabled MLWP systems to achieve skill levels comparable to, and in some cases exceeding, state-of-the-art numerical models for short- to medium-range forecasts (Lang et al., 2025). Perhaps overlooked is the fact that most ML models are indirectly blending data science with physics-based modelling: although they do not simulate physical processes directly, most rely heavily on training data generated by physics-based models - particularly reanalysis products that encapsulate decades of physically consistent simulations.

ML models offer transformative opportunities: orders-of-magnitude reductions in computational cost to produce a forecast, the ability to generate large ensembles at low expense, the possibility to meet new customer needs, and the potential for faster update cycles and higher-fidelity forecasts without super-linear increases[1] in energy consumption. These come at a time when the growing cost of traditional high-performance computing is making it more difficult and expensive to improve physics-based model performance at the rate required to continue the quiet revolution.

Despite these opportunities, current MLWP approaches carry significant risks and uncertainties. Their reliance on historical data raises concerns about robustness under climate change, extreme events, and data-sparse regions. Interpretability and trust remain major challenges, particularly for operational use where transparency and physical consistency are critical. There are similar and arguably even greater challenges for climate, where we are looking to predict a future unseen climate state. For this there is likely a need to augment historical data with physical climate model simulations that provide training data for future climates. Key issues are around the ability of ML models to capture extremes and out-of-sample events (Sun et al, 2025), interpretability (O'Loughlin et al., 2025) and reliability. In addition, the practical challenges of transitioning large, complex, and highly tuned physics-based forecasting systems - delivering a wide range of products - pose significant barriers to the *wholesale* adoption of ML models. It is also currently the case that challenges remain around the temporal resolution that is needed for useful weather forecasts (most MLWP models provide 6-hourly data). With these challenges in mind it is the case that (with a small number of exceptions) most MLWP models remain "research models". Some of these obstacles can, however, be bypassed in the near term through blended approaches that combine ML components with existing physics-based systems, offering a more incremental and manageable path to innovation. It is likely that blended approaches will be needed in the long-term for climate, where we are looking to enhance and not replace physical climate model simulations.

*Direct-from-Observations (DOP) Methods*

Before discussing blended methods, it is also worth commenting on Direct-from-Observations (Direct Observation Prediction or DOP) approaches, which are emerging and aim to train ML models exclusively on observational data; bypassing the dependence on physics-based simulations. These methods promise to reduce inherited biases from numerical models and can in principle bring in more and better observations to improve future skill without a dependence on a (potentially slow and expensive) model development cycle. DOP application can be used for the generation of analyses, but can also be used for end-to-end predictions. However, DOP methods face significant challenges, including sparse observational coverage, variations in the type and accuracy of available data, and the difficulty of capturing the full state of the Earth system without model-based assimilation. Observing systems are themselves costly and complex, and moving to prediction systems that depend exclusively on observational input streams would introduce new vulnerabilities, particularly regarding robustness to interruptions, calibration changes, or upgrades in the observing network.

---

[1] Typically physics-based models improve accuracy by increasing resolution – a 2-fold increase in horizontal resolution requires roughly 8 times the compute power due to the numerical conditioning and therefore reduction of the model timestep.

DOP approaches represent an important research frontier and a potential disruptor in future model development and deployment strategies. If reliable DOP systems do come along, it is likely that caution will be given in their immediate adoption and so blended implementation strategies would be sensible for operational deployment.

## 3. Blending

The term *blending* is already well-established in the weather and climate community to describe the combination of multiple physics-based models within an ensemble (e.g. Rust et al., 2023). These blends may incorporate outputs from models of different origins or resolutions. Here, we argue for the extension of the use of *blended modelling* to encompass a broader range of approaches that combine machine learning and/or physics-based methods in weather and climate prediction. Given the diverse ways in which ML and physics-based methods can be combined, we further identify sub-categories within this expanded definition of blending. To support clearer communication and strategic planning, we advocate for a typology that distinguishes five main categories: *independent-physics-based*, *hybrid-integrated-ML*, *hybrid-composite-ML*, *augmented-ML*, and *independent-ML*. This is captured in the schematic in Fig. 1.

### Independent-Physics-Based

At one end of our typology we have established physics-based methods. This category represents the well-established methods that have underpinned our computer-based weather forecasting and climate prediction and projection capabilities since their emergence more than 60 years ago. They represent the culmination of years of scientific research in understanding and modelling the physical processes controlling our weather patterns and climate system. They have proven to provide reliable guidance and have seen continuous improvement in their skill through the introduction of new techniques, better observational data and increasing computer power. However, a significant drawback is their relatively significant computational cost.

The majority of operational weather forecasts and climate projections are currently based on independent-physics-based models and modelling systems, although these can often use a blend of outputs from different physics-based models.

### HYBRID-INTEGRATED-ML

Within our typology, hybrid methods are considered to be those that use ML and physics-based techniques that run alongside one another to deliver a predicted state. We further refine this category into hybrid-integrated-ML and hybrid-composite-ML(described below). The key distinction is the way in which the two methods are coupled together. For hybrid-integrated methods there is a tight 2-way coupling such that the physics-based component influences the ML-based component and vice-versa. One example here would be the emulation of physical parametrizations using a neural net. This ML-based component is built into the physics-based model and so it both inputs physical model state into the neural net and receives updated parametrized increments (e.g. Morcrette, 2025). Another example would be a system-aware approach that trains a neural network as a source term - representing sub-grid parametrizations and/or corrections to truncation errors - embedded within the physics-based model core, allowing the learning process to propagate through the entire dynamical system (e.g., Kochkov et al., 2024).

Hybrid-integrated methods offer two main potential advantages. First, they could represent a higher level of complexity within the neural network than a physics-based approach could achieve at the same computational cost. Second, the neural network may have a significantly lower computational cost compared to an equivalent physics-based system. Considerations and motivations around computational cost also need to factor in the implications on and from (hybrid) compute architectures. Some hybrid methods may also provide a lower barrier to adoption in complex technical workflows providing an incremental evolution of large operational systems.

### HYBRID-COMPOSITE-ML

Unlike hybrid-integrated methods, hybrid-composite methods only require a one-way flow of information from the ML component to the physics-based component or vice versa. This means that the components can be run asynchronously. An example in this category is that of the nudging technique described by Husain et al., 2025. In this example, a MLWP model is first used to provide a prediction of the large-scale weather evolution. This prediction is then read into the physics-based forecast model and the predicted fields (in this case temperature and winds) are used to nudge the physical-model evolution and guide the large-scale state. While the overhead of the computational cost of the MLWP model is relatively low, the improved accuracy of the large-scale and long lead time evolution can give significant improvements in forecast skill.

Beyond nudging, other examples in this category could come from using output from one method applied to a global forecast to provide driving boundary conditions for a regional forecast using the other method. It could also encapsulate ML learned error correction in the data assimilation or through the forecast.

### AUGMENTED-ML

This category consists of ML methods applied or MLWP added at a post-processing step. The defining feature is that an ML technique operates on the output of a physics-based or hybrid system to produce new or augmented data. Examples include adding ML-based model solutions to increase the number of members in a physics-based ensemble, creating blended outputs from multiple models. These models could have similar or different characteristics – leading to something akin to a super-ensemble (if characteristics are similar) or a multi-model ensemble (if characteristics are different). Another example might be generating new or enhanced diagnostic products or parameters using the output from physics-based or hybrid models.

INDEPENDENT-ML

At the far end of our typology, this category represents the move to use methods that are entirely based on data driven ML approaches. While the training of these systems may rely on data generated from physics-based or hybrid models (e.g. reanalysis data or climate model simulations), deployment does not require any use of the traditional techniques associated with physics-based modelling. A key advantage of these techniques is the significant reduction in computational cost, reducing power consumption, reducing time to solution and making them accessible to a broader user base.

## 4. Benefits of blending

*Benefits, Opportunities and Challenges of different blending approaches*

**Accuracy & Quality:** Accuracy and quality of a weather forecast or climate projection are the key considerations when developing a blended prediction system. Improvements in accuracy and quality can arise from the development of models with reduced errors and biases, more comprehensive representation of physical processes, larger ensembles that better capture uncertainty, and better quality, quantity and use of observations. All the blending categories offer examples of approaches whereby one or more of these improvements can be made. However, there are some specific considerations to note:

- Independent-physics or hybrid solutions more than likely require significant computational cost to deliver significant gains in accuracy.
- Independent-ML solutions have demonstrated a significant improvement in the skill of large-scale, synoptic weather metrics. This is also true of some integrated-ML approaches such as system-aware (physics-informed) learning (Kochkov et al, 2024).
- The accuracy of fine scale features of MLWP models is less well proven and while much work has been done recently to improve this (e.g., Subich et al, 2025) there remains some uncertainty about whether regional or global emulators can accurately capture local, high impact phenomena (Kendon et al, 2025).
- Blends that can combine the large-scale synoptic accuracy of ML-based methods with fine-scale physics-based methods may prove to be the optimal approach for weather forecasting (there is currently no consensus on the potential of this for climate).

- There remains some uncertainty about the accuracy of ML-based models when presented with scenarios outside of their training data (e.g., grey swan events, Sun et al, 2025). Augmented ensemble blends may provide one way to help mitigate this risk that accompanies ML-based model adoption.
- For blends where computational expense can be reduced, ensemble size can be increased and so lead to improved solution quality.

**Computational speed & cost:** The primary driver for the adoption of ML-based methods - beyond potential improvements in accuracy - comes from the huge potential for reductions in computational cost. This translates directly into cost of power consumption, money or time-to-solution. While significant upfront costs are needed for training of ML-based models, these are substantially smaller than the ongoing operational costs of deploying a physics-based model. When it comes to the inference step of a ML-based model, there are reductions in:

- Overall energy consumption - translating directly into reduced carbon footprint and cutting financial costs associated with power usage.
- Time to solution - leading to a faster turnaround of long climate integrations or time-critical weather forecasts.
- Per-member cost of an ensemble - allowing for larger and more diverse ensembles.
- Hardware resource requirements for running the model - enabling researchers and institutions (particularly those in developing nations) who lack access to advanced HPC facilities.

It is important to recognize that while hybrid and augmented systems facilitate incremental adoption of ML-based methods, they may also incur additional costs because they require running both an ML-based system and a physics-based system in parallel. Some of this cost is associated with additional resources for development and maintenance (which may double), while a key driver of this cost is also the difference in hardware requirements: ML models typically run most efficiently on GPU or TPU architectures, while historically physics-based models have relied primarily on CPUs. This mismatch can lead to increased infrastructure complexity and expense. However, advances in hybrid architectures (Fusco et al., 2024, Schieffer et al. 2024) and performance-portable physics-based models (Adams et al., 2019, Bauer et al., 2021, Ubbiali et al., 2025 ) are expected to reduce these overheads over time.

**Trust & explainability:** Trust and explainability are essential for maintaining confidence in weather forecasts and climate projections. This is particularly true for critical services such as severe weather warnings, emergency response, and adaptation and mitigation climate policy (this is of paramount importance when providing information for future unseen climates). Physics-based models are considered to inherently provide a high level of trust because they are grounded in well-understood scientific principles and offer comprehensive diagnostic outputs that can be interrogated to confirm self-consistency and adherence to those principles. In contrast, ML-based systems pose challenges for explainability due to their data-driven nature. Today, physics-based research into explainability (O'Loughlin et al. 2025) is essential to deepen our understanding of how ML-models capture and reproduce physical laws and in so doing enhance transparency in the decision-making processes.

Hybrid and augmented approaches are seen as a possible route to balance concerns about trust and explainability. They offer a foundation of the scientific rigor of the physics-based system which maintains physical consistency and provide opportunities for cross-validation against established evaluation frameworks and principles.

**Agility & flexibility:** Agility and flexibility are critical for ensuring that prediction systems can adapt to evolving scientific knowledge, emerging technologies, and changing user needs. Pragmatic choices concerning operational deployment must be considered alongside the science when building a new system, if it is to be useful in the short term. Longer term ambitions - perhaps moving to radically different delivery systems - must be sufficiently adaptable to mitigate the risk of failure and to "lock-in" to a single solution.

Agility of a system reflects the ability to rapidly adapt. An agile system has the potential to support faster update cycles, more frequent forecasts, and tailored products for diverse applications, from local weather services to climate risk assessments. Importantly, these approaches democratize access to advanced prediction capabilities by reducing computational barriers, making high-quality predictions feasible for research groups and institutions with limited resources.

Various blending options help to provide a pathway for such systems to be more flexible, adaptable and agile. Hybrid-integrated methods that build on existing operational physics-based models will have the advantage of a smooth transition into operations - not requiring new technical infrastructure or downstream processing. Hybrid-composite methods and augmented approaches introduce new requirements to support the ML-based components, but offer a route to implementation of more agile ML-based solutions.

It is also the case that ML approaches are readily customised to deliver a prediction meeting a specific user requirement, making product development more agile. This aspect could also deliver better quality bespoke products.

Taken together, the range of blending strategies highlights both the opportunity and the responsibility that comes with integrating ML and physics-based prediction systems. Blended approaches allow us to exploit the complementary strengths of each paradigm, while offering practical routes to improved accuracy, reduced cost, and greater agility. Yet it is also clear that several of the more ambitious concepts, including fully independent ML-based or Direct-from-Observations systems, remain scientifically and operationally unproven. Their feasibility will depend on overcoming challenges around data coverage, robustness, extrapolation skill, and the integration of new scientific knowledge. For this reason, blended and hybrid pathways provide not only a promising avenue for near-term gains but also a pragmatic hedge against uncertainty, enabling the prediction enterprise to innovate rapidly while ensuring that essential levels of trust, explainability, and resilience are preserved as the field continues to evolve.

## 5. Examples of applications of different blending types

Blended modelling is already being explored across several active projects at the Met Office and the wider weather and climate communities. We briefly present two case studies to illustrate how different blending strategies – nudging for weather forecasting and downscaling for climate - are being applied to enhance forecast skill, operational agility, and allow more rapid scientific investigation.

**Forecast Adjustment (Nudging)**

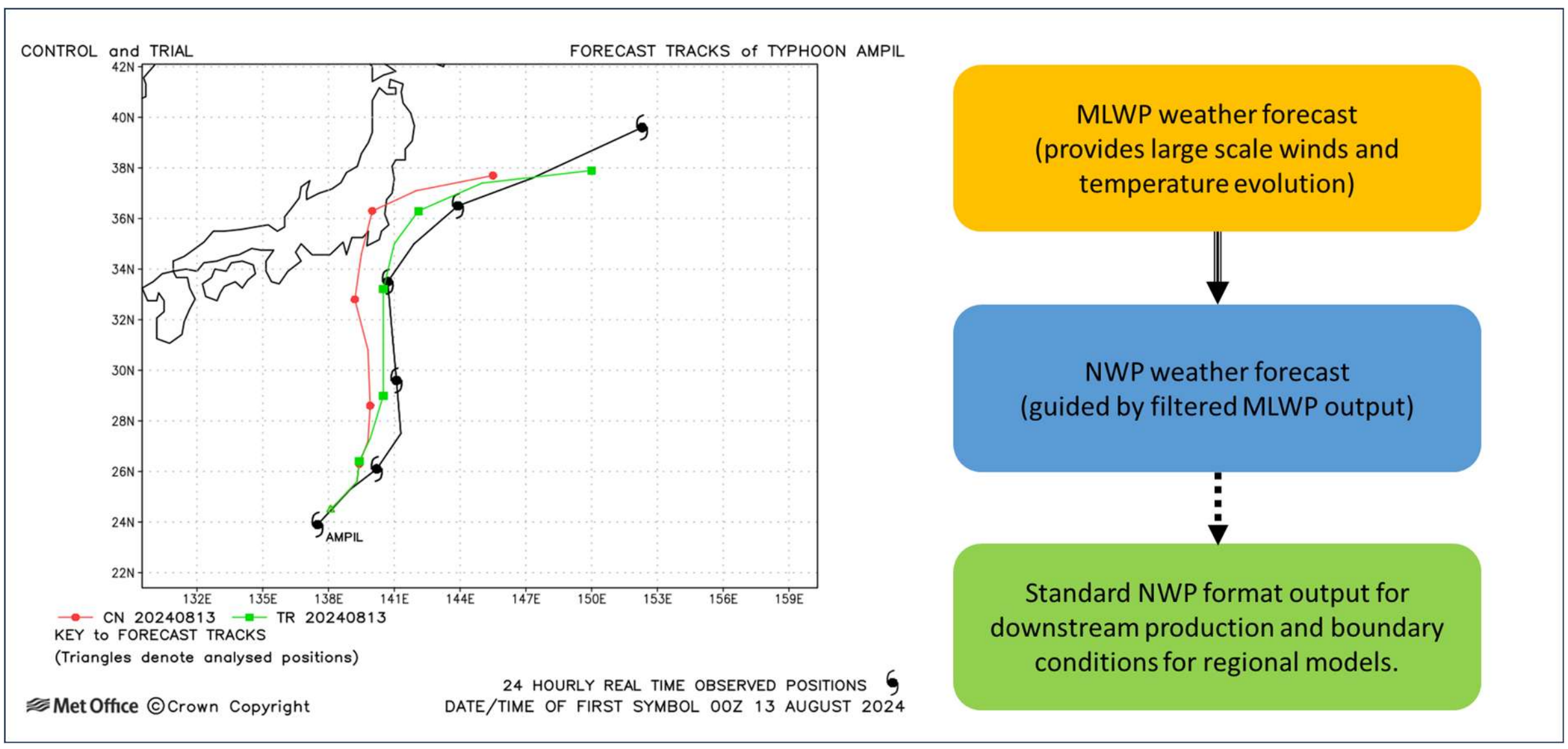

Fig. 2. Forecast tracks for Typhoon Ampil, showing NWP-only solution (red), hybrid-composite solution (green) and observed track (black). The large-scale guidance from the MLWP creates a better forecast of the typhoon track, whilst the small-scale accuracy of the NWP ensures typhoon structure and intensity measures, such as minimum sea-level pressure or maximum wind-speed, are unaffected.

One promising approach under investigation is the use of ML-based forecasts to nudge a physics-based model during runtime (Husain et al., 2025). This technique (Fig 2.) aims to improve the large-scale evolution of the forecast while retaining trusted features and smaller-scale accuracy. In this hybrid-composite configuration, the ML weather-prediction (MLWP) system is run separately and in advance of the NWP forecast to produce predictions of the broad synoptic pattern, which are then ingested by the physics-based model. These fields act as gentle corrections, steering the NWP forecast toward the more skilful large-scale trajectory suggested by the ML model, without replacing or interfering with the physical model's detailed representation of atmospheric processes. This same concept has also been extended to a probabilistic ensemble framework, combining IFS-ENS with a machine-learned ensemble and improving large-scale skill and tropical cyclone track performance without degrading ensemble spread (Polichtchouk et al, 2026).

Nudging offers an agile and highly practical route for introducing the benefits of MLWP into operations without disrupting existing forecasting systems. The physics-based NWP model still provides the same output fields that it would without the MLWP steering and so delivery of downstream products (for which MLWP models are not yet mature enough to deliver) continues unimpeded.

**Physics-Model-Driven Downscaling**

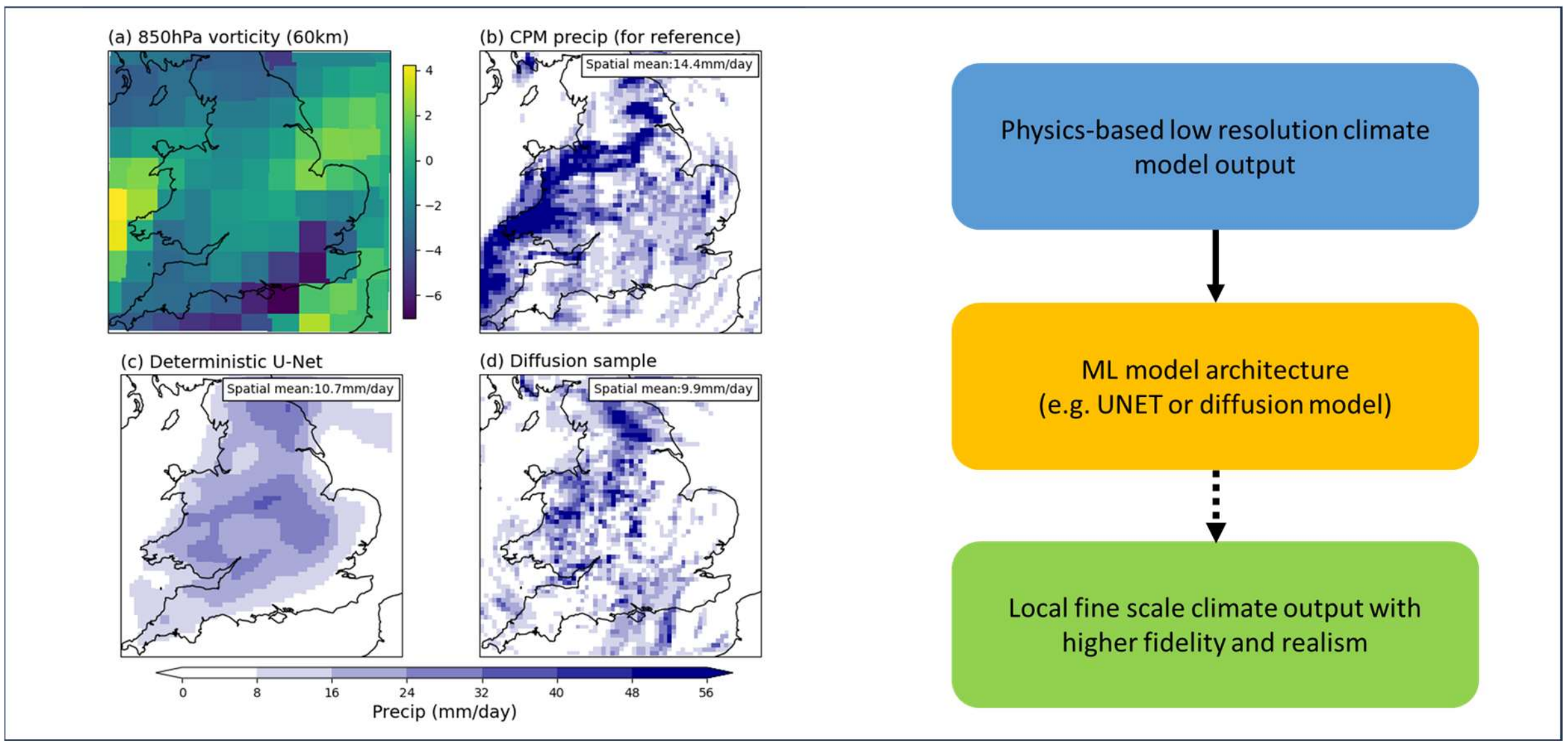

Fig. 3. AI downscaling emulators, trained on existing Regional Model simulations, provide a cost effective and rapid way to translate global weather and climate information to higher resolution locally relevant information. This example illustrates a Deterministic (UNET) and Generative (Diffusion) ML approach that predicts the regional model rainfall, based on large scale synoptic weather drivers (temperatures, winds and/or vorticity, humidity and pressure). The Generative models produce stochastic samples of the local weather consistent with these large scale drivers, providing a potential pathway to sample estimates of uncertainty in local impacts. These stochastic samples better capture structure and intensities of spatially heterogenous fields, such as rainfall.

Physics-model-driven downscaling (Fig 3.) maps naturally onto the Hybrid-Composite-ML category in the typology. In the domain of regional climate modelling, this approach is showing real potential for adding value in the provision of local climate change information (Kendon et al, 2025). These methods predict a local-scale variable such as precipitation or temperature from large-scale conditions ('predictors') describing the state of the atmosphere. While current applications involve obtaining the large-scale predictors from a physics-based global climate model, the methodology could equally make use of a data-driven climate model as confidence and experience grows in these. Physics-based regional climate models crucially are used for training and evaluation of the ML downscaling emulator, and thus this approach enhances but does not replace physics-based modelling.

The reduction in computational expense of ML-downscaling compared to physics-based regional climate simulations opens up the possibility of large ensembles or very long simulations at a fraction of the cost. At the Met Office we are exploring the potential for these methods to deliver new capability for (1) km-scale event attribution; (2) convection-permitting climate projections by downscaling a larger sample of global climate models and thus more comprehensively sampling uncertainties at local scales; (3) local detail for UNSEEN (an approach using large model ensembles to characterise unprecedented extremes); (4) downscaling monthly-to-decadal predictions; and (5) urban scale modelling. Key science challenges remain, however, before these ML-downscaling emulators can be used in earnest in climate services. These include performance for extremes and out-of-sample events, challenges around transferability to out-of-sample global climate models, multi-variate downscaling, memory and sub-daily temporal coherence, and evaluating information from generative samples (Kendon et al, 2025).

## 5. Conclusion

Blended modelling represents a pragmatic, scalable, and strategically important pathway for the future of weather and climate prediction. The rapid development of machine learning methods offers huge opportunities to advance scientific, operational and societal needs. ML is not a replacement, but something that we should look to exploit alongside trusted and established physics-based systems, ensuring we combine the strengths of both to deliver more capable, efficient, and reliable prediction services.

There are myriad different ways in which these approaches can be used alongside one another. The ‘optimal’ way to blend these approaches will depend on target use cases, system maturity and complexity, and will naturally evolve as time goes by and our understanding and trust of emerging capabilities matures. This paper aims to provide a framework for discussing and identifying different ways of combining ML and physics-based models and prediction systems through a typology. It further describes some of the benefits that may be found in combining ML and physics-based systems alongside one another.

As the landscape continues to evolve, the value of a clear and shared language becomes even more important. By distinguishing the different ways ML and physics-based systems can interact, this typology helps to articulate ambitions, assess trade-offs, and prioritise investment in a structured way. It also highlights that no single approach will meet all needs: different blends offer different balances of accuracy, computational efficiency, agility, trust, and operational practicality. This typology is intended to support that evolution with clarity and consistency.

*Acknowledgments*

We thank Julian Heming for producing the forecast track of Typhoon Ampil used in Fig. 2. We are also grateful to the many colleagues and collaborators who contributed through discussions that helped shape the ideas and framing of this paper.

*Data Availability Statement.*

No new datasets.